# Calibration-free wavelength modulation spectroscopy based on even-order harmonics

**YIHONG WANG,**[1] **BIN ZHOU,**[1,*] **AND CHANG LIU**[2,*]

[1]*School of Energy and Environment, Southeast University, Nanjing 210096, China*
[2]*School of Engineering, The University of Edinburgh, Edinburgh EH9 3JL, U.K*
*\*zhoubinde@seu.edu.cn, c.liu@ed.ac.uk*

**Abstract:** This paper proposes a novel and rapid calibration-free wavelength modulation spectroscopy algorithm based on even-order harmonics. The proposed algorithm, analytically deduced from Voigt line-shape function, only involves simple algebraic operations to describe the actual gas absorption spectra, thus eliminating the time-consuming simulations and line-shape fitting procedures adopted in traditional algorithms. Instead of acquiring the entirely scanned absorption line-shape, the proposed technique only requires extraction of the peak values of the harmonics. This characteristic significantly benefits gas diagnosis at elevated pressure and/or temperature, in which the entirely scanned absorption is very difficult to be obtained due to the broadened line-shapes. The proposed algorithm is validated by both numerical simulation and condition-controlled experiment, indicating millisecond-level calculation of gas parameters with the relative error less than 4% in the experiments.

## 1. Introduction

Tunable diode laser absorption spectroscopy (TDLAS) has been widely employed to measure gas properties, such as temperature and species concentration, given its advantages of contactless, rapid, and high sensitivity [1-5]. As a modality of the TDLAS technique, wavelength modulation spectroscopy (WMS) can achieve very high sensitivity with strong robustness to background noise, and therefore, has been extensively applied in harsh environments with, for example, strong turbulence, high pressure and/or temperature. Although WMS technique offers many benefits, quantitative WMS measurement is challenging since harmonic signals that are used for calculating the gas parameters strongly depend on the absorption line-shape. Consequently, the measurement results generally need to be calibrated with a known gas mixture, which is impractical in case of (a) poorly-known gas conditions in industrial applications, (b) difficulty in deploying a well-controlled calibration equipment, in particular, with high-temperature and high-pressure gases, and (c) optical inconsistency, e.g. transmission of the optical windows, between the calibration equipment and the target reactors. In the past decade, the above-mentioned challenges have stimulated the development of multiple calibration-free WMS methods. Implementation of the calibration-free WMS can be mainly categorized as model-dependent and model-independent methods.

The model-dependent method is to calculate the gas parameters using the analytical model of WMS, alongside the well-characterized laser parameters. Li et al. [6] pioneered the development of the model-dependent method by simulating WMS signals as a function of laser parameters and absorption spectra. To eliminate the influence of light intensity fluctuations, Rieker et al. [7] normalized the second-order harmonic signal of WMS by the first-order harmonic signal, i.e. WMS-2f/1f, and further included laser-specific tuning characteristics in the spectral-absorption model to infer gas properties. However, the retrieval accuracy of gas parameters using these methods highly depends on comprehensive characterization of both the line-shape parameters, e.g. collisional broadening coefficients, of the target transition, and the laser parameters, e.g. amplitudes and phases of laser intensity modulation and frequency modulation. In practice, these parameters measured in the laboratory can be very different from those measured on-site, due to the complex gas composition and wide ranges of temperature and pressure.

In contrast, the model-independent method, achieved by waveform fitting of the experimentally measured WMS-2f/1f (or WMS-nf/1f) signal with a simulated one, is more suitable for practical applications since less strict or even no requirement is placed on characterizing the line-shape parameters and laser parameters. This advantage significantly speeds up its application towards in situ diagnosis of reactive flows in harsh environments, for example, at elevated pressure and/or temperature [8-10], as the retrieval is independent of the prior information of line broadening. However, the waveform fitting requires time-consuming iterations until the residual is converged to an acceptable level. When the target absorption is interfered by nearby transitions, the result of the multi-parameter fitting cannot be always unique, as the iterations can be converged to the local optimum instead of the global one. To address these issues, Chen et al. [11] recently proposed to use the peak values of zero-, second-, and fourth-order harmonics, enabling the direct and fast measurement without complex simulations and curve fittings. However, Chen's algorithm is derived from the Lorentzian line-shape function, which is not applicable in low-pressure scenarios. Furthermore, the zero-order harmonic signal is ease of noise contamination, leading to reduced fidelity when the measurement is deployed at the presence of environmental noise owing to mechanical vibration, beam steering and window fouling.

Although higher order harmonics have lower signal strength [12, 13], they can benefitted from larger signal-to-background ratios (SBRs) [14-17] and stronger robustness at the extreme conditions where absorption spectra are significantly broadened and large modulation index is required. In this work, we propose, for the first time, a calibration-free WMS algorithm based on even-order harmonics (order $\geq$ 2). The amplitude of each harmonic signal, used for retrieving the gas properties, is analytically derived from the Voigt line-shape function, enabling not only speedy measurement down to milliseconds, but also a general suitability for various degrees of line-shape broadening. The rest of the paper is organized as follows: Section 2 details the fundamentals of the proposed algorithm. Numerical simulation and proof-of-concept experiment are carried out to validate the proposed algorithm in Section 3 and Section 4, respectively. The paper is finally summarized in Section 5.

## 2. Methodology

### 2.1 Theoretical background

In this subsection, some basics are presented to facilitate the derivation of the proposed method. WMS is implemented by imposing a high-frequency sinusoidal modulation with angular frequency $\omega$ on a slowly varying diode laser injection current. As a result, the instantaneous laser frequency $v$ and the laser intensity $I_t$ can be expressed as:

$$v = v_0 + a\cos(\omega t), \qquad (1)$$

$$I_t = \bar{I}_0 (1 + \sum_{k=1}^{\infty} i_k \cos(k\omega t + \psi_k)), \qquad (2)$$

where $v_0$ is the laser center frequency, $a$ the modulation depth. $\bar{I}_0$ is the averaged laser intensity. $i_k$ is the $k$-th order Fourier coefficient of the laser intensity. $\psi_k$ is the $k$-th order phase shift between intensity modulation and frequency modulation. For an isolated absorption line, the modulation index $m$ is defined by $2a/\lambda$, where $\lambda$ represents the full width at half maximum (FWHM) of the absorption line.

According to the Beer-Lambert law, the spectroscopic absorbance can be written as:

$$\alpha(v) = -\ln(I_t / I_0) = A\varphi(v), \qquad (3)$$

where $I_t$ and $I_0$ are the transmitted and incident laser intensities, respectively. $A$ is the integral absorption area. $\varphi(v)$ is the line-shape function, described by Voigt line-shape function, is a convolution of Lorentzian and Gaussian line-shape functions, noted as $\varphi_L(v)$ and $\varphi_G(v)$, respectively. $\varphi(v)$ can be approximated by [18]

$$\varphi(v) = c_L\varphi_L(v) + c_G\varphi_G(v), \quad (4)$$

$$\varphi_L(v) = 2/(\pi\lambda)[1+((v-v_0)/(\lambda/2))^2]^{-1}, \quad (5)$$

$$\varphi_G(v) = 2/\lambda\sqrt{\ln 2/\pi}\exp[-\ln 2((v-v_0)/(\lambda/2))^2], \quad (6)$$

where $c_L$ and $c_G$ are the weights of the Lorentzian and Gaussian broadening coefficients, $\lambda_L$ and $\lambda_G$, respectively. $c_L$, $c_G$ and $\lambda$ can be calculated by

$$c_L = 0.6818817 + 0.6129331d - 0.1838439d^2 - 0.1156844d^3, \quad (7)$$

$$c_G = 0.3246017 - 0.6182531d + 0.1768139d^2 + 0.1210944d^3, \quad (8)$$

$$\lambda = 0.5346\lambda_L + \sqrt{0.2166\lambda_L^2 + \lambda_G^2}, \quad (9)$$

where the line-shape parameter $d$ is defined as

$$d = (\lambda_L - \lambda_G)/(\lambda_L + \lambda_G). \quad (10)$$

A dimensionless parameter $\Delta = 2(v - v_0)/\lambda$ is introduced to describe the relative offset of the laser center frequency, and therefore, $\varphi(v)$ can be reformulated as

$$\varphi(v) = \frac{2}{\pi\lambda}[c_L L(\Delta, m) + c_G\sqrt{\pi\ln 2}G(\Delta, m)], \quad (11)$$

where $L(\Delta, m)$ and $G(\Delta, m)$ are the time-dependent peak normalized Lorentzian and Gaussian linear functions respectively, and are expressed as

$$L(\Delta, m) = [1+(\Delta+m\cos(\omega t))^2]^{-1}, \quad (12)$$

$$G(\Delta, m) = \exp[-\ln 2(\Delta+m\cos(\omega t))^2]. \quad (13)$$

Following Arndt's model [14, 19], the analytical expressions of the $n$-th harmonic of $L(\Delta, m)$ can be written as

$$L_n(\Delta) = \frac{\varepsilon_n i^n}{2m^n}\frac{[\sqrt{(1-i\Delta)^2+m^2}-(1-i\Delta)]^n}{\sqrt{(1-i\Delta)^2+m^2}} + c.c., \quad (14)$$

where $c.c.$ denotes the conjugate complex number of the preceding term. $\varepsilon_0 = 1$ and $\varepsilon_n = 2$ for $n \geqslant 1$. Given no offset of the laser center frequency, i.e., $\Delta = 0$, the amplitude of each harmonic obtained from Eq. (14) can be analytically expressed by

$$h_n^L = \begin{cases} \frac{\varepsilon_n}{m^n}\frac{(\sqrt{1+m^2}-1)^n}{\sqrt{1+m^2}} & \text{for even } n \\ 0 & \text{for odd } n \end{cases}. \quad (15)$$

By using infinite series theory [20], the amplitude of each harmonic of $G(\Delta, m)$ can be analytically expressed by

$$h_n^G = \begin{cases} 2\exp(-m^2\ln 2/2)I_{n/2}(m^2\ln 2/2) & \text{for even } n \\ 0 & \text{for odd } n \end{cases}, \quad (16)$$

where $I_{n/2}(z) = i^{n/2}J_{n/2}(iz)$. $J_{n/2}$ is the first kind of modified Bessel functions of order $n/2$.

As indicated by Eq. (11), $\varphi(v)$ is a linear function of $L(\Delta, m)$ and $G(\Delta, m)$. When the linearity of Fourier transform is employed, the amplitude of the $n$-th harmonic $h_n$ of the given the spectral absorbance $\alpha(v)$ can be expressed by the following linear combination:

$$h_n = \begin{cases} 4A/\pi\lambda c_L p_n(m) + 4A/\pi\lambda c_G q_n(m) & \text{for even } n \\ 0 & \text{for odd } n \end{cases}, \quad (17)$$

where $p_n(m)$ and $q_n(m)$ are defined as

$$p_n(m) = \frac{\varepsilon_n}{2m^n}\frac{(\sqrt{1+m^2}-1)^n}{\sqrt{1+m^2}}, \quad (18)$$

and
$$q_n(m) = \sqrt{\pi \ln 2} \exp(-m^2 \ln 2/2) I_{n/2}(m^2 \ln 2/2), \quad (19)$$

respectively. Dependence of the amplitudes of $p_n(m)$ and $q_n(m)$ on the modulation index $m$ is shown in Fig. 1. For a given $m$, both $p_n(m)$ and $q_n(m)$ decrease as the even harmonic order $n$ increases. Consequently, the first several nonzero even harmonics should be adopted to achieve higher signal-to-noise ratios (SNRs) of the WMS measurement.

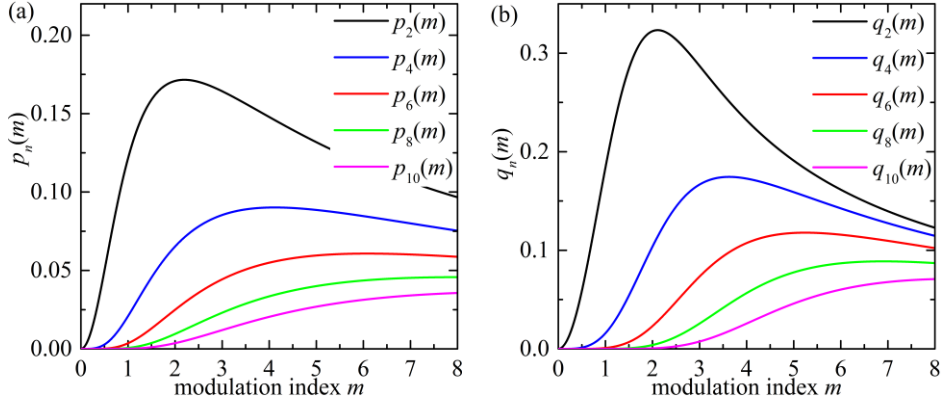

**Fig. 1.** Dependence of (a) $p_n(m)$ and (b) $q_n(m)$ on the modulation index $m$.

### 2.2 Fundamentals of the proposed algorithm

The purpose of the proposed algorithm is to calculate the spectral parameters from the measured amplitudes of nonzero even harmonics. As indicated by Eq. (17), the amplitudes of the first $N$ nonzero even harmonics adopted can be rewritten as:

$$\mathbf{H} = k_p \mathbf{P}(m) + k_q \mathbf{Q}(m), \quad (20)$$

where $\mathbf{H} \in \mathbb{R}^N$ is the vector of measured harmonics amplitudes, defined as $\mathbf{H} = (h_2, h_4, \ldots, h_{2N})^T$. $\mathbf{P}(m) \in \mathbb{R}^N$ and $\mathbf{Q}(m) \in \mathbb{R}^N$ are $m$-dependent basis vectors, defined as $\mathbf{P}(m) = (p_2(m), p_4(m), \ldots, p_{2N}(m))^T$ and $\mathbf{Q}(m) = (q_2(m), q_4(m), \ldots, q_{2N}(m))^T$, respectively. The coefficients $k_p$ and $k_q$ to be solved are defined as $k_p = 4Ac_L/(\pi\lambda)$ and $k_q = 4Ac_G/(\pi\lambda)$, respectively. Eq. (20) shows that the vector $\mathbf{H}$ is an element of the column space in the matrix $\mathbf{M} = (\mathbf{P}(m), \mathbf{Q}(m))$. Consequently, $\mathbf{H}$ and its projection, noted as $\mathbf{H}'$, should coincide strictly:

$$\mathbf{H} - \mathbf{H}' = (\mathbf{E} - \mathbf{M}(\mathbf{M}^T\mathbf{M})^{-1}\mathbf{M}^T)\mathbf{H} = \mathbf{0}, \quad (21)$$

where $\mathbf{E}$ is the $N$-dimensional identity matrix. Theoretically, with $\mathbf{H}$ in hand, the modulation index $m$ can be calculated from the nonlinear equation (21). However, the measurement noise deviates the results of $\mathbf{H} - \mathbf{H}'$ in Eq. (21) from its desired value of $\mathbf{0}$. In this work, the modulation index $m^*$ is calculated by solving the following minimization problem:

$$m^* = \arg\min_m \| (\mathbf{E} - \mathbf{M}(\mathbf{M}^T\mathbf{M})^{-1}\mathbf{M}^T)\mathbf{H} \|_2. \quad (22)$$

When $N = 2$, $\mathbf{M} \in \mathbb{R}^{2\times2}$ is non-singular matrix, and thus the projection matrix, $\mathbf{M}(\mathbf{M}^T\mathbf{M})^{-1}\mathbf{M}^T$, is reduced to the identity matrix, $\mathbf{E}$. In this case, $\mathbf{H} - \mathbf{H}'$ always equals to $\mathbf{0}$, and arbitrary modulation index $m^*$ can be chose. To make the above minimization problem solvable, the dimension of $\mathbf{H}$ must be no less than 3.

With $m^*$ in hand, $k_p$ and $k_q$ in Eq. (20) can be calculated by using the least square method:

$$\begin{bmatrix} k_p \\ k_q \end{bmatrix} = (\mathbf{M}^{*T}\mathbf{M}^*)^{-1}\mathbf{M}^{*T}\mathbf{H}, \quad (23)$$

where $\mathbf{M}^* = (\mathbf{P}(m^*), \mathbf{Q}(m^*))$. $k_p/k_q$ is given by:

$$\frac{k_p}{k_q} = \frac{c_L}{c_G} = \frac{0.6818817 + 0.6129331d - 0.1838439d^2 - 0.1156844d^3}{0.3246017 - 0.6182531d + 0.1768139d^2 + 0.1210944d^3}. \quad (24)$$

Obviously, $k_p/k_q$ is only determined by the line-shape parameter $d$. Fig. 2 shows $k_p/k_q$ is monotonically dependent on $d$ when $-0.85 \leq d \leq 0.85$. That is to say, for any given $d$, there is a unique value of $k_p/k_q$. It is worth mentioning the line-shape is described as the Voigt function when $-0.85 \leq d \leq 0.85$. The line-shape functions can be characterized by the Lorentzian and the Gaussian functions for $0.85 < d \leq 1$ and $-1 \leq d < -0.85$, respectively, resulting into simplified calculation of $k_p/k_q$ to be detailed in subsection 2.3.

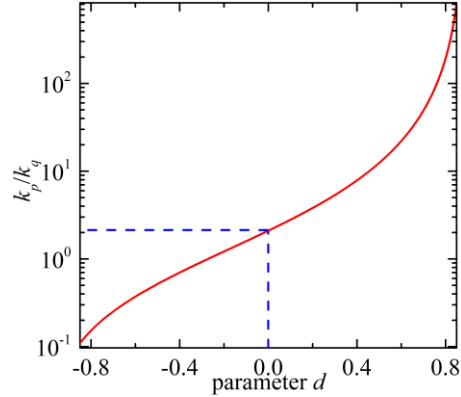

**Fig. 2.** Dependence of $k_p/k_q$ on $d$ when $-0.85 \leq d \leq 0.85$.

Therefore, the FWHM of the Voigt line-shape can be calculated by
$$\lambda = 2a / m^*, \tag{25}$$
and thus the integrated absorbance area can be calculated by
$$A = \pi \lambda k_p / (4c_L) \text{ or } A = \pi \lambda k_q / (4c_G). \tag{26}$$
The integrated absorbance areas obtained from two individual transitions can be used to infer the gas temperature using two-line thermometry [21, 22]. With the temperature in hand, the gas concentration can be calculated from one of the integrated absorbance areas. The flow chart for calculating gas properties using the proposed algorithm is shown in Fig. 3.

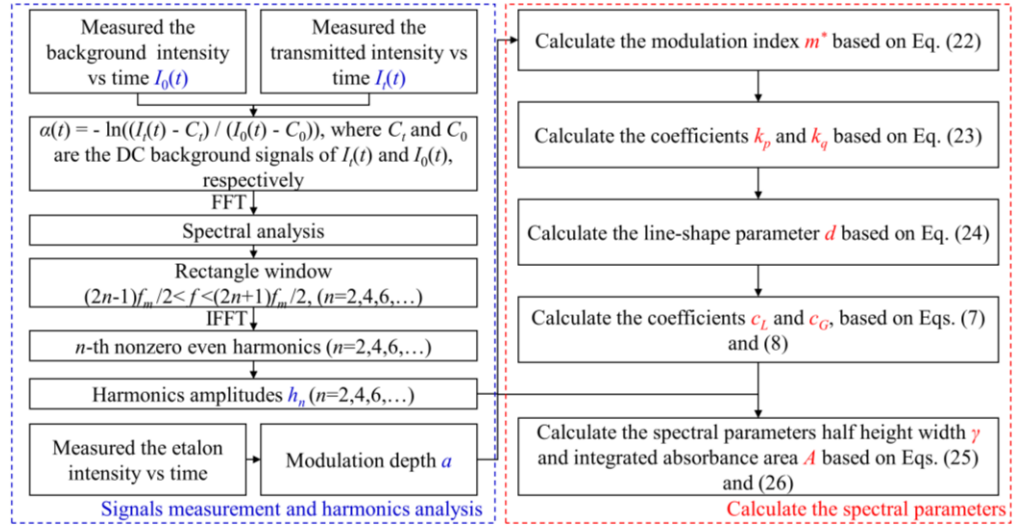

**Fig. 3.** Flow chart for calculating gas properties using the proposed algorithm.

*2.3 Simplified version for Lorentzian and Gaussian line-shape functions*

As noted above, the line-shape function is Lorentzian when $0.85 < d \leq 1$, while that is Gaussian when $-1 \leq d < -0.85$. In these two cases, Eq. (20) can be reformulated as

$$\mathbf{H} = \begin{cases} k_p \mathbf{P}(m) & \text{for } 0.85 < d \leq 1, \\ k_q \mathbf{Q}(m) & \text{for } -1 \leq d < -0.85, \end{cases} \tag{27}$$

where $k_p = k_q = 4A/(\pi\lambda)$. Obviously, $\mathbf{H}$ is linear-dependent on $\mathbf{P}(m)$ or $\mathbf{Q}(m)$. Thus, the modulation index $m^*$ can be solved from the following minimization problems:

$$m^* = \begin{cases} \arg\min_m -\dfrac{\mathbf{H} \cdot \mathbf{P}(m)}{\|\mathbf{P}(m)\|_2} & \text{for } 0.85 < d \leq 1, \\ \arg\min_m -\dfrac{\mathbf{H} \cdot \mathbf{Q}(m)}{\|\mathbf{Q}(m)\|_2} & \text{for } -1 \leq d < -0.85. \end{cases} \tag{28}$$

Similarly, to make Eq. (28) solvable, the dimension of $\mathbf{H}$ must be no less than 2. Substituting $m^*$ into Eq. (27), $k_p$ and $k_q$ can be obtained as follows:

$$k_p = \frac{\|\mathbf{H}\|_2}{\|\mathbf{P}(m^*)\|_2}, \tag{29}$$

$$k_q = \frac{\|\mathbf{H}\|_2}{\|\mathbf{Q}(m^*)\|_2}. \tag{30}$$

With $k_p$ or $k_q$ in hand, the line-shape parameters, such as FWHM, integrated absorbance area, and thus the gas parameters can be obtained in the same way described in subsection 2.2.

### 2.4 Parameter optimization

Eq. (20) indicates $\mathbf{H}$ can be represented linearly by $\mathbf{P}(m)$ and $\mathbf{Q}(m)$. The coefficients $k_p$ and $k_q$, therefore, are critical to solve the gas properties. According to the theory of linear algebra, the numerical stability for solving $k_p$ and $k_q$ can be characterized by the condition number of $\mathbf{M}^T\mathbf{M}$. In case of an inverse problem with larger condition numbers, a small perturbation in $\mathbf{H}$ will cause large perturbations in $k_p$ and $k_q$. Therefore, it is necessary to choose an appropriate modulation index $m^*$ to make $k_p$ and $k_q$ noise-insensitive.

Fig. 4 shows the dependence of the condition number of $\mathbf{M}^T\mathbf{M}$ on modulation index $m$. It can be seen the condition number decreases as $m$ starts to increase from 0, and reaches a local minimum value when $m$ is around 1.5. Then, condition number increases and tops when $m$ is around 3. After that, the condition number monotonically decreases as $m$ increases. When $m \geq 3.5$, the condition number is below the previous local minimum value, indicating the ever-stronger numerical stability for solving $k_p$ and $k_q$. However, as indicated in Fig. 1, an excessively large $m$ leads to the reduction of harmonic amplitude. Therefore, the optimal $m$ is selected to be 3.5.

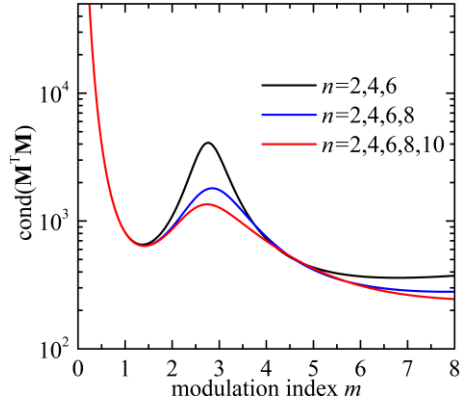

**Fig. 4.** Dependence of the condition number of $\mathbf{M}^T\mathbf{M}$ on modulation index $m$.

### 3. Numerical simulation

*3.1 Simulation setup*

The proposed algorithm is numerically validated for temperature measurement using water vapor ($H_2O$) transitions. The absorption transitions at 7185.60 cm$^{-1}$ and 7444.36 cm$^{-1}$, that are widely used in two-line thermometry [21, 22], are employed in both the simulation and experiment afterwards. In the simulation, the atmospheric-pressure gas concentration is set as 0.05. The absorption length is 20 cm. The gas temperature ranges from 500 K to 2,500 K. The modulation depths for the transitions at 7185.60 and 7444.36 cm$^{-1}$ are set to 0.131 cm$^{-1}$ and 0.127 cm$^{-1}$, respectively. The scanning depths for the both transitions are set to 0.15 cm$^{-1}$. The frequencies of sinusoidal scan and modulation in the simulation are 100 Hz and 10 kHz, respectively.

According to the HITRAN database [23], there are several neighbor transitions, with parameters detailed in Table 1, interfering the target transitions. Although each of the transitions appeared as a single absorption line-shape, the combined absorbance shown in Fig. 5 exhibits overlap of absorption line-shapes of nearby transitions with different lower energy levels. The overlap is particularly significant for the transition at 7444.36 cm$^{-1}$, covering two interference transitions spaced up to 0.0133 cm$^{-1}$. As shown in Fig. 5 (b), the two interference transitions result in significant asymmetry of the target absorbance with additional broadening of the line shape.

**Table 1. Parameters of the selected transitions at around 7185.60 cm$^{-1}$ and 7444.36 cm$^{-1}$. For each transition, the table shows its wavenumbers, line strengths ($S(T_0)$) at $T_0$ = 296 K, pressure broadening-induced half-widths for itself ($\xi_{self}$) and air ($\xi_{air}$), energies of the lower-state ($E''$), and coefficients of temperature dependent broadening ($n_{air}$) [23].**

| line index | wavenumber (cm$^{-1}$) | $S(T_0)$ (cm$^{-2}$atm$^{-1}$) | $\xi_{self}$ (cm$^{-1}$atm$^{-1}$) | $\xi_{air}$ (cm$^{-1}$atm$^{-1}$) | $E''$ (cm$^{-1}$) | $n_{air}$ |
|---|---|---|---|---|---|---|
| 1 | 7185.596571 | 0.00490296 | 0.371 | 0.0342 | 1045.0583 | 0.63 |
|   | 7185.597300 | 0.01470888 | 0.195 | 0.0413 | 1045.0578 | 0.65 |
|   | 7185.394270 (neighbor lines) | 0.00127968 | 0.435 | 0.0924 | 447.25230 | 0.63 |
|   | 7185.400770 (neighbor lines) | 0.0002375592 | 0.315 | 0.0780 | 1474.9803 | 0.89 |
| 2 | 7444.351800 | 0.00054064 | 0.366 | 0.0199 | 1774.7503 | 0.44 |
|   | 7444.368159 | 0.0001539336 | 0.250 | 0.0188 | 1806.6701 | 0.41 |
|   | 7444.370600 | 0.000462024 | 0.194 | 0.0153 | 1806.6694 | 0.41 |
|   | 7444.562600 (neighbor lines) | 0.000155496 | 0.189 | 0.0209 | 1774.6153 | 0.44 |
|   | 7444.695400 (neighbor lines) | 0.0005456 | 0.358 | 0.0534 | 1437.9678 | 0.51 |

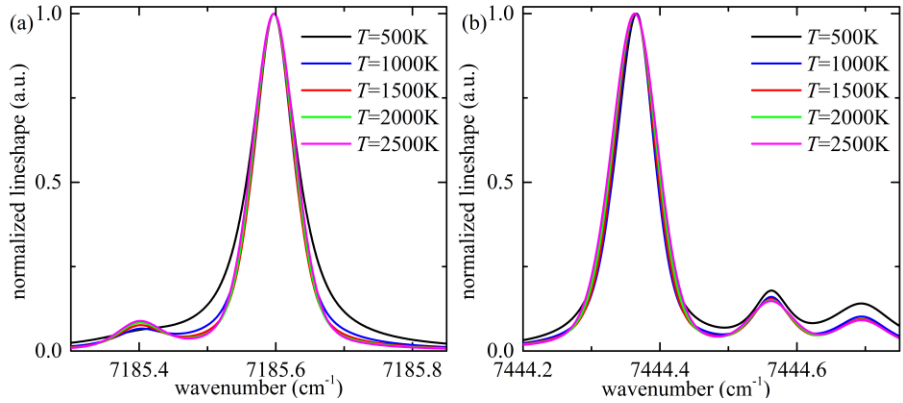

**Fig. 5.** The normalized line-shapes at different temperatures for the transitions at around (a) 7185.60 cm$^{-1}$ and (b) 7444.36 cm$^{-1}$, respectively.

With the above-mentioned settings of the WMS simulation, the modulated absorbance signal at the temperature of 1000 K for the two target transitions are obtained, and thus the corresponding second- to tenth-order even harmonics are acquired by the FFT spectral analysis [24]. As shown in Fig. 6, the higher-harmonics give lower amplitude, but have the same order of intensities with the lower-order ones. In the proposed algorithm, only the peak values of the even harmonics are needed, as illustrated in Eqs. (20) – (30), thus significantly simplifying and fastening of data processing in comparison with the traditional waveform fitting algorithm.

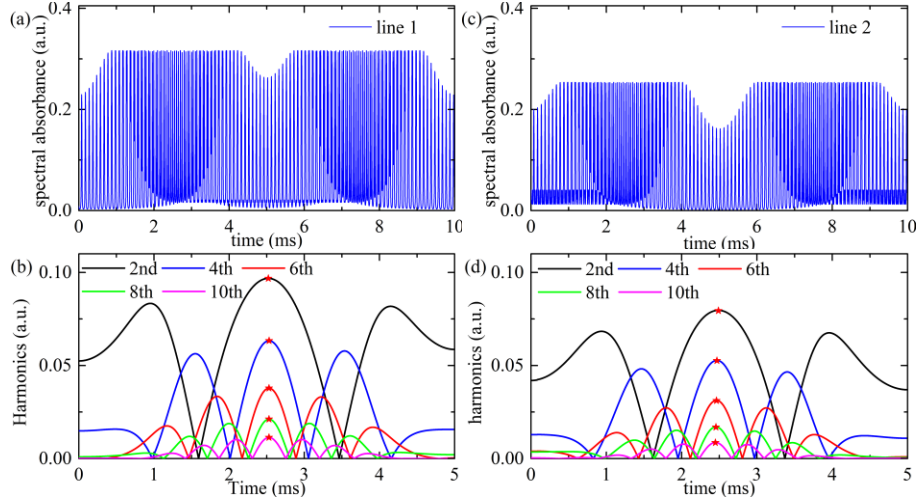

**Fig. 6.** Simulated WMS raw signal and the extracted even harmonics acquired by the FFT spectral analysis at 1000 K. (a) and (b) show the modulated absorbance signal for the transitions at around 7185.60 cm$^{-1}$ and 7444.36 cm$^{-1}$ with the simulated H$_2$O spectral absorption, respectively. (c) and (d) shows the second- to tenth-order even harmonics extracted from (a) and (b), respectively.

*3.2 Calculation of the modulation index*

As noted in Eq. (22), the modulation index $m^*$ should be firstly solved by minimizing the objective function $\mathbf{F} = \|(\mathbf{E}-\mathbf{M}(\mathbf{M}^T\mathbf{M})^{-1}\mathbf{M}^T)\mathbf{H}\|_2$, which depends on the selected orders of the even harmonics. For the two H$_2$O transitions at 7185.60 cm$^{-1}$ and 7444.36 cm$^{-1}$, Fig. 7 shows the dependence of $\mathbf{F}$ on the modulation index $m$ in three cases with different combinations of the $n$th order even harmonics, i.e. $n$ = 2, 4, 6 (case 1), $n$ = 2, 4, 6, 8 (case 2), and $n$ = 2, 4, 6, 8, 10 (case 3), respectively. For both the transitions, $\mathbf{F}$ vs. $m$ variation tendencies are same for all the three cases, with the minimum $\mathbf{F}$ obtained with similar $m$ values. For the transitions at 7185.60 cm$^{-1}$, the minimum $\mathbf{F}$ is obtained when $m$ equals to 3.843, 3.866, and 3.872 for cases 1, 2 and 3, respectively. For the transitions at 7444.36 cm$^{-1}$, the corresponding $m$ equals to 3.796, 3.814, and 3.815 for cases 1, 2 and 3, respectively.

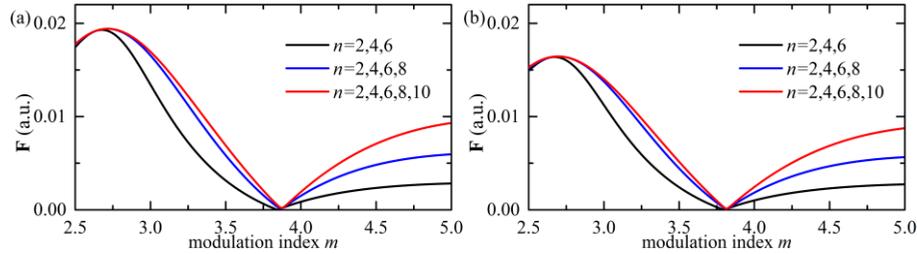

**Fig. 7.** Dependence of the objective function $\mathbf{F}$ on the modulation index $m$ at 1000 K for the transitions at (a) 7185.60 cm$^{-1}$ and (b) 7444.36 cm$^{-1}$.

*3.3 Simulation results with noise-free and noise-contaminated data*

With the modulation index $m^*$ and the amplitudes of the even-order harmonics in hand, we can calculate the integrated absorbance areas for the two transitions according to Eq. (26), and their ratio to infer the gas temperature. We will evaluate retrieved temperature using the proposed algorithm with noise-free and noise-contaminated measurements, respectively.

For the above three cases, Fig. 8 shows the relative error of the retrieved temperature using noise-free data. As the target transitions are interfered by the nearby transitions, minor errors exist in the retrieved temperature even if no noise is imposed on the measurement. For all the three cases, the relative errors are less than 0.5% in the temperature range of 500 - 1250 K. When the temperature is above 1250 K, the retrievals degenerate in all the three cases, with the errors in cases 1 and 2 less significant than that in case 3. The worst scenarios at 2500 K give the relative errors of -2.135%, -2.406%, and -3.175% for cases 1, 2, and 3, respectively.

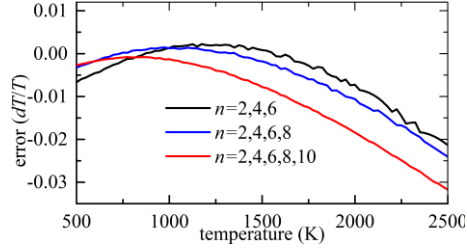

**Fig. 8.** Relative errors of the retrieved temperature with different harmonic combinations.

Then, white noise is imposed on both the transmitted and incident laser intensities, i.e. $I_t$ and $I_0$, according to the following formula:

$$\alpha'(v) = -\ln(\frac{I_0 \exp(-\alpha(v)) + noise_t}{I_0 + noise_0}), \quad (31)$$

where $noise_t$ and $noise_0$ are white noise generated by the *randn* function of MATLAB with a mean value of 0 and a standard deviation of 0.01% × average laser intensity, equivalent to a signal-to-noise ratio (SNR) of 60 dB. This noise level is dominated by the thermal noise from the detector, signal amplification, and digitization circuits [25]. To characterize the noise performance of the proposed algorithm, the average relative error and the relative standard deviation (RSD) of the temperature retrieved from 100 repetitive simulations are shown in Figs. 9 (a) and (b), respectively. The average relative error shown in Fig. 9 (a) indicates similar results as those using the noise-free data. As the temperature sensitivities of the chosen $H_2O$ transitions decrease as the temperature increases [22, 26], the retrieved temperature suffers from more significant fluctuations, indicated by the larger values of RSDs shown in Fig. 9 (b), at higher temperatures. In addition, involvement of more harmonics contributes to lower RSDs, and thus stronger robustness of the proposed algorithm.

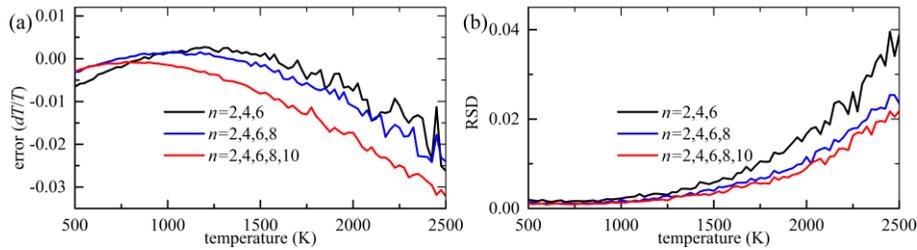

**Fig. 9.** Performance evaluation of the proposed algorithm with noise-contaminated data. (a) and (b) show the average relative error and relative standard deviation (RSD) of the temperature retrieved from 100 repetitive simulations.

## 4. Experimental setup and results

### 4.1 Experimental setup

In this section, experiments, with setup shown in Fig. 10, were carried out to validate the proposed algorithm. The quartz optical cell has three zones, all fitted with wedged windows to suppress etalon interference. The two outer zones were purged with dry air to prevent unwanted

$H_2O$ absorption along the optical path, while the center zone, with the path length of 20 cm, was filled with a uniform gas mixture at a controlled temperature between 773 and 1273 K. The gas mixture is obtained from the dry air going through $H_2O$ liquid, and therefore, contains the target $H_2O$ molecules for the absorption measurement. Three K-type thermocouples were equally spaced along the optical cell to measure the temperature of the target gas. The optics, detector and the intermediate open paths were purged with $N_2$ to remove interfering $H_2O$ absorption in the room air.

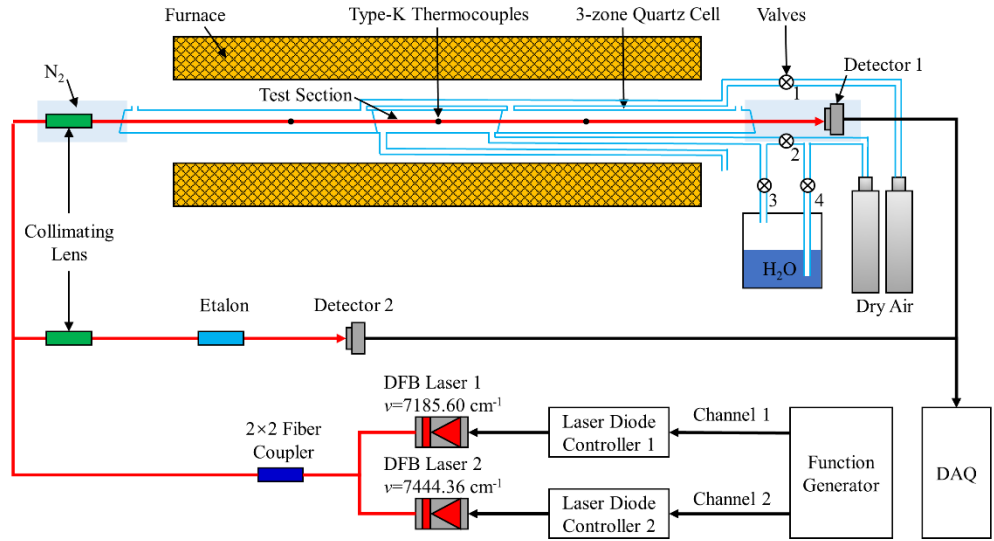

**Fig. 10.** Schematic diagram of the experimental system.

Two distributed feedback (DFB) diode lasers operating near 7185.60 cm$^{-1}$ (NTT, NLK1E5GAAA) and 7444.36 cm$^{-1}$ (NTT, NLK1B5EAAA) were used to probe the $H_2O$ transitions. Each laser was controlled independently by a laser controller (Stanford Research Systems, LDC501), which adjusts the temperature and the injection current of the laser. To achieve wavelength modulation, the injection current was driven by a function generator (RIGOL, DG1062Z) which provided a 100 Hz sinusoidal scan signal superimposed with a 10 kHz sinusoidal modulation signal. The two lasers were operated in time division multiplexing (TDM) mode [27]. To monitor DC background signals such as the dark current of the detector and thermal radiation, there is a 10 ms non-laser interval between the neighboring wavelength scan of the two lasers. The two lasers were combined and then split by a 2×2 single-mode fiber-coupler, with one output being guided through an etalon with a free spectral range (FSR) of 0.01 cm$^{-1}$ for wavelength monitoring and the other output penetrating the target gas cell. Each laser beam was detected by a photodetector (Thorlabs, PDA10CS-EC) and further sampled by a 14-bit data acquisition card (NI Corporation, PXIe-5170R). The digitization rates for sampling the etalon signal and the absorption signal are 25 Mega samples per second (Msps) and 1 Msps, respectively.

The optical cell was firstly heated to the desired temperature and filled with dry air to obtain the absorption-free transmission. Then, the center zone was evacuated and filled with the $H_2O$/air mixture. When the adsorption and transient effects was stabilized, the absorption transmissions were recorded at the given furnace temperature. The temperature of the optical cell was then adjusted to repeat the entire measurement process. In the experiment, the flow rates for both the purging gas and the target gas mixture were small and stabilized to reduce the influence of convective heat transfer on the temperature set by the furnace. Therefore, two mass flow meters were used to control both gas flow rates at 100 standard cubic centimeter per minute (sccm). Since the optical cell is open to the air, the gas pressure inside the cell is same as the atmospheric pressure, that is 0.998 atm, measured by the pressure gauge (MKS, type660).

### 4.2. Experimental results and discussions

Figs. 11 (a) and (b) shows the sampled background transmission $I_0(t)$ and the absorption transmission $I_t(t)$ at 973 K, respectively. Both $I_0(t)$ and $I_t(t)$ are averaged for 100 repetitive measurements to reduce the interference of measurement noise. Although the background transmission is not an ideal absorption-free signal, caused by the small purging flow of the dry air, logarithmic transformation in Eq. (3) neutralizes any impact of the background absorption on the temperature retrieval. In addition, the rising edges of both the second periods of $I_0(t)$ and $I_t(t)$ are adopted in the harmonic analysis in the experiment, since the first period, just after switching the lasers, is under stabilization. Subsequently, the even-order harmonics demodulated by FFT spectral analysis are shown in Figs. 12 (a) and (b) for the transitions at 7185.60 cm$^{-1}$ and 7444.36 cm$^{-1}$, respectively. Finally, the amplitude of each harmonic is extracted by conic fitting of the partial waveform nearby the peak.

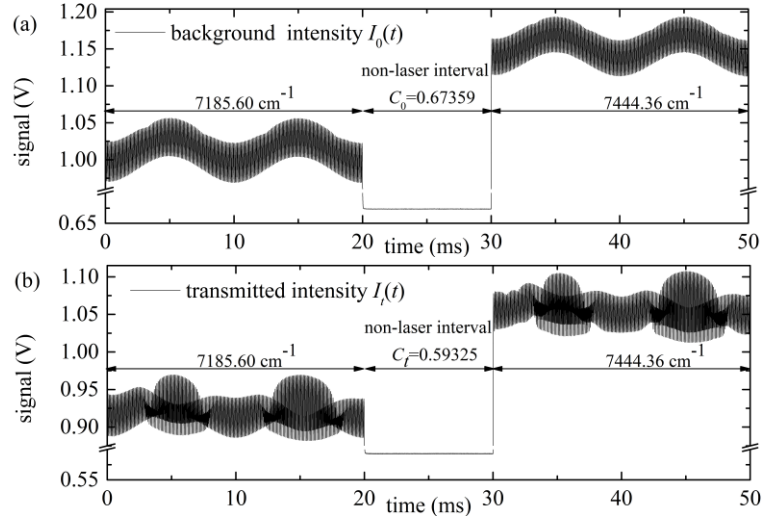

**Fig. 11.** The experimentally sampled (a) background intensity and (b) transmitted intensity signals at 973 K.

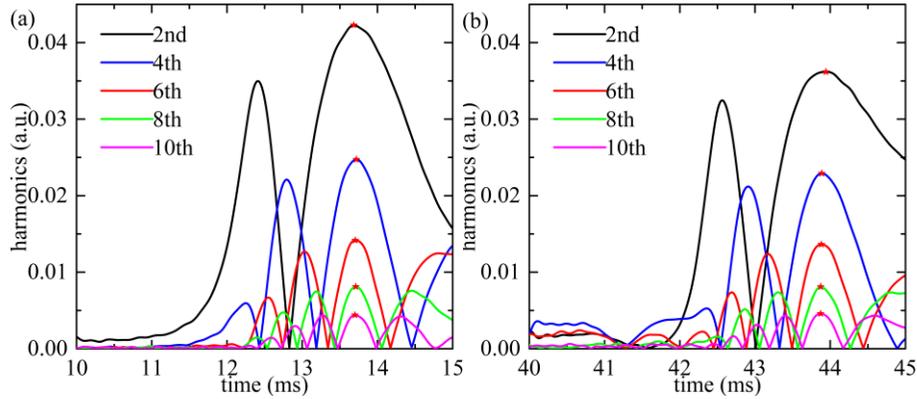

**Fig. 12.** The even-order (2$^{nd}$ to 10$^{th}$) harmonics demodulated by FFT spectral analysis for the transitions at (a) 7185.60 cm$^{-1}$ and (b) 7444.36 cm$^{-1}$.

With the optimal modulation index determined in subsection 3.2 and the extracted amplitudes of the even-order harmonics, the temperature is retrieved using the proposed algorithm by following Eqs. (20)-(30). To examine the performance of the proposed algorithm, the retrieved temperature values are compared with those measured by the thermocouple in the range of 773 - 1273 K with an interval of 100 K. Similar as the simulation, the three cases with different combinations of the even-order harmonics are verified in the experiment. As shown in Fig. 13 (a), the temperature retrieved in all the three cases agree well with that measured by the thermocouple. The maximum relative errors, characterized in Fig. 13 (b), are 2.59%, 3.60% and 2.82% for cases 1, 2 and 3, respectively.

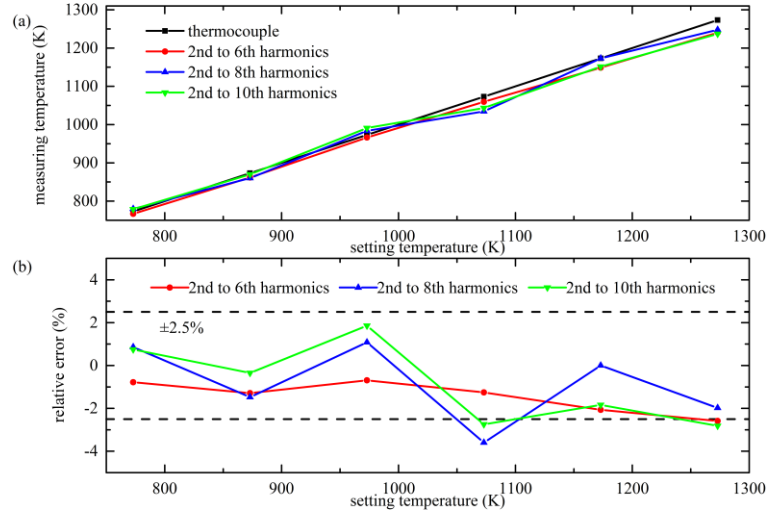

**Fig. 13.** Comparison of the temperature measured using the proposed algorithm and the thermocouple. (a) and (b) show the absolute measured temperature and the relative error, respectively.

To evaluate the superiority of the proposed algorithm in terms of computational cost, the waveform fitting algorithm [28] was also employed to analyze the same acquired data. Fig. 14 shows the measured and fitted 1f-normalized WMS-2f (WMS-2f/1f) waveforms with the fitting residuals for the transitions of 7185.60 cm$^{-1}$ and 7444.36 cm$^{-1}$ at 973 K. The waveform fitting is implemented using the samples on the wavenumber free from the interference transitions. Therefore, 4000 samples on the left wing of absorbance are used for fitting for the transition 7185.60 cm$^{-1}$ (Fig. 14 (a)), while 2800 samples of the right wing are used for fitting for the transition 7444.36 cm$^{-1}$.

It is worth noting the WMS-2f/1f waveform fitting algorithm is sensitive to the initial value of parameters to be fitted. Improper initial values may increase the computational time, and even make the solution fall into the local optimum. In this work, the furnace temperature given by thermocouples and other needed line-shape parameters retrieved by the proposed algorithm were taken as the initial values of the fitting algorithm. In contrast, the proposed algorithm is always converged with no restriction attached to the initial value. All the computational process in this section were conducted by a computer with an Intel Core i7-1165G7 @ 2.80GHz CPU.

The computational cost results for these two algorithms are summarized in Table 2. The proposed algorithm in the temperature range of 773 - 1273 K gives the average computational time of 0.137 s, 0.142 s, and 0.144 s for cases 1, 2, and 3, respectively. Although the appropriate initial values were given, the average computational time of the WMS-2f/1f waveform fitting algorithm still reaches the average time of 22.902 s. That is to say, the computational efficiency of the proposed algorithm is improved by at least two orders of magnitude.

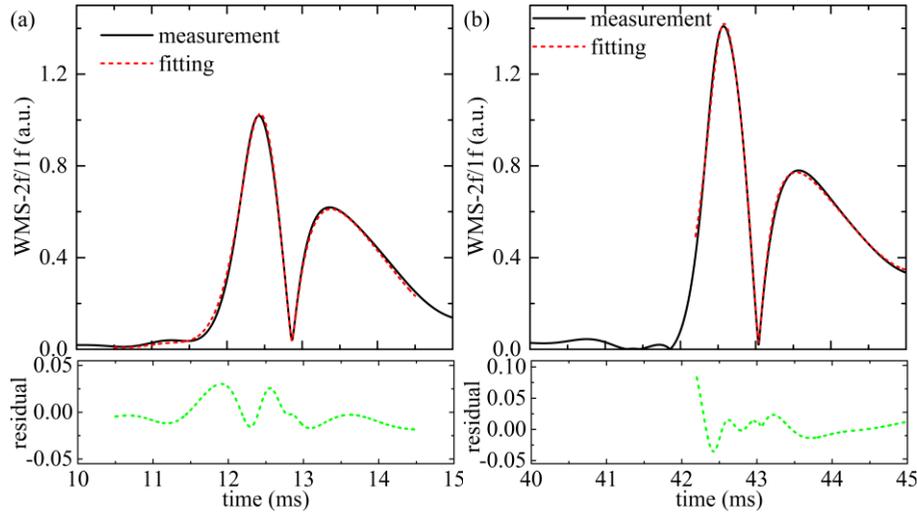

**Fig. 14.** Measured and fitted WMS-2f/1f waveform for the transitions at (a) 7185.60 cm$^{-1}$ and (b) 7444.36 cm$^{-1}$ with the residuals shown in above subplots.

**Table 2. Summary of computational time with different algorithms.**

| algorithm | WMS-2f/1f waveform fitting [28] | the proposed algorithm in this work | | |
|---|---|---|---|---|
| | | case 1 | case 2 | case 3 |
| average computational time (s) | 22.902 | 0.137 | 0.142 | 0.144 |
| speed up | / | 167.168 | 161.282 | 159.042 |

## 5. Conclusions

A novel calibration-free WMS algorithm, analytically deduced from the Voigt line-shape function, is proposed for accurate retrieval of gas properties using the amplitudes of the even-order harmonics. As only the peak values of the harmonics are needed, the proposed algorithm substantially benefits WMS measurement with large line-shape broadening, for example, measurements in high-pressure and high-temperature shock-wave wind tunnels. For the two pre-selected $H_2O$ transitions, numerical simulation was firstly carried out to determine the optimal modulation index for different combinations of the even-order harmonics, and examine their accuracy and noise resistance. Simulation results indicate the relative errors are less than 0.5% in the temperature range of 500 - 1250 K and stronger robustness when more harmonics are involved. Furthermore, experimental validation of proposed algorithm shows a maximum relative error of 3.6% in the temperature range of 773 - 1273 K. Compared with the traditional WMS-2f/1f waveform fitting algorithm, the computational efficiency achieved by the proposed algorithm is improved by at least two orders of magnitude.

**Funding.** National Key Research and Development Program of China (2017YFB0603204); Global Challenges Research Fund (LMIC Travel and Partnerships Fund).

**Disclosures.** The authors declare no conflicts of interest.

**Data availability.** Data underlying the results presented in this paper are not publicly available at this time but may be obtained from the authors upon reasonable request.